\documentclass{article}
\usepackage[dvips]{graphicx}
\begin{document}
\title{Scattering and Trapping of Nonlinear Schr\"odinger Solitons in External Potentials}
\author{Hidetsugu Sakaguchi and Mitsuaki Tamura\\
Department of Applied Science for Electronics and Materials,\\ Interdisciplinary Graduate School of Engineering Sciences,\\
 Kyushu University, Kasuga, Fukuoka 816-8580, Japan}
\maketitle
{\bf abstract}\\
Soliton motion in some external potentials is studied using the nonlinear Schr\"odinger equation. Solitons are scattered by a potential wall. Solitons propagate almost freely or are trapped in a periodic potential. The critical kinetic energy for reflection and trapping is evaluated approximately with a variational method.\\
\\
keywords: soliton, nonlinear Schr\"odinger equation, Lagrangian
\\
\\

The nonlinear Schr\"odinger equation has been intensively studied 
as a soliton equation, which is derived as an envelope equation by several perturbation methods~\cite{Karp}. Solitons in optical fibers are typical ones described by the nonlinear Schr\"odinger equation~\cite{Hase}. Recent successful observation of solitons in  Bose-Einstein condensates (BECs) raises the theoretical consideration of this equation again~\cite{Khay}. The equation is called the Gross-Pitaevskii equation in the research field of BECs. In the BECs, matter waves obey the nonlinear Schr\"odinger equation with external potentials. For example, external magnetic fields are applied to confine the BECs. Interference patterns of laser beams generate external periodic potentials. Solitons in external potentials such as harmonic potential, periodic potential and stepwise potential have been studied by several authors~\cite{Moura,Frau,Baiz}. 
But solitons in a potential wall or potential well were not studied in detail. 
In this letter, we will study the scattering and trapping of solitons by a potential wall (or well) and a periodic potential. In particular, we focus on soliton behaviors like classical mechanical particles, and evaluate the critical energies for the transmission and the trapping. 

The nonlinear Schr\"odinger equation with a potential term is written as 
\begin{equation}
i\frac{\partial \phi}{\partial t}=-\frac{1}{2}\frac{\partial^2\phi}{\partial x^2}-|\phi|^2\phi+U(x)\phi,
\end{equation}
where $\phi(x,t)$ is the wave function, $U(x)$ is an external potential. 
Without the nonlinear term, this equation is equivalent to the Schr\"odinger equation with mass $m=1$ and $\hbar=1$. When $U(x)=0$, the nonlinear Schr\"odinger equation has a soliton solution
\begin{equation}
\phi(x,t)=A{\rm sech}\{A(x-vt)\}e^{ip(x-vt)-i\omega t},
\end{equation}
where $v=p$ and $\omega=-A^2/2+v^2/2$. 

Firstly, we investigate the nonlinear Schr\"odinger equation in a simple rectangular potential, that is,
\begin{equation}
U(x)=U_0,\;\;{\rm for}\;\;x_0<x<x_0+d,\;\;\;
U(x)=0,\;\;{\rm for }\;\; x<x_0, \;x>x_0+d.
\end{equation}
If $U_0$ is positive (negative), $U(x)$ represents a potential wall (well).
We consider first the case of $U_0>0$. 
If the nonlinear term is absent in eq.~(1), this is a typical potential problem of the quantum mechanics. The transmission coefficient $T$ of matter waves is expressed as 
\begin{equation}
T=\left [1+\frac{U_0^2\sin^2(\alpha d)}{4E(E-U_0)}\right ]^{-1},\;{\rm for}\;\;E>U_0,\;\;\;T=\left [1+\frac{U_0^2\sinh^2(\alpha^{\prime} d)}{4E(E-U_0)}\right ]^{-1},\;{\rm for}\;\;E<U_0,
\end{equation}
where $E$ denotes the kinetic energy $E=v^2/2$, and $\alpha$ and $\alpha^{\prime}$ are defined as  $\alpha=\sqrt{2(E-U_0)}$ (for $E>U_0$) and $\alpha^{\prime}=\sqrt{2(U_0-E)}$ (for $E<U_0$).  We study numerically the scattering of a soliton by the potential wall.  Figure 1(a) displays the time evolution of a soliton passing through a potential wall with $U_0=0.5$ and $d=1$. The system size is $L=100$ and the potential wall locates at $x=x_0=L/2=50$. The initial velocity of the soliton is $v=0.75$ and the amplitude is $A=1$. The soliton is split into two pulses by the potential wall. The transmission coefficient is 0.527, which is calculated as $T=\int_{L/2}^L|\phi|^2dx/\int_0^L|\phi|^2dx$ after the scattering. Figure 1(b) displays the transmission coefficient as a function of the initial velocity $v$. The dashed curve is the transmission coefficient eq.~(4).  The transmission coefficient of a soliton is close to the coefficient eq.~(4) for $v>0.9$, which might imply that the effect of the nonlinear term becomes relatively weaker as the kinetic energy is increased. The transmission coefficient decreases more sharply than the quantum mechanical particle as $v$ is decreased. 
\begin{figure}[t]
\begin{center}
\includegraphics[width=10cm]{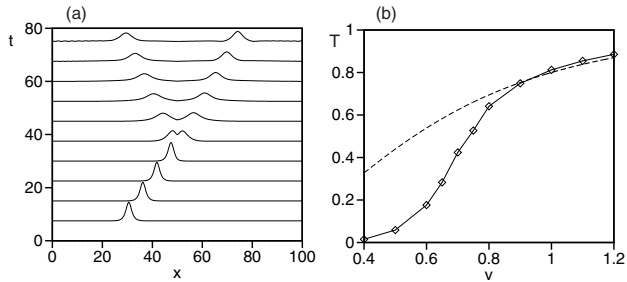}
\end{center}
\caption{(a) Time evolution of $|\phi(x,t)|$ for $U_0=0.5,A=1$ and $v=0.75$.
A soliton is splitted into two pulses by the potential wall of $d=1$ located at $x=L/2=50$. (b) Transmission coefficient $T$ as a function of the initial velocity $v$. The dashed curve is the transmission coefficient for the quantum mechanics expressed by eq.~(4)}
\label{f1}
\end{figure}

\begin{figure}[t]
\begin{center}
\includegraphics[width=14cm]{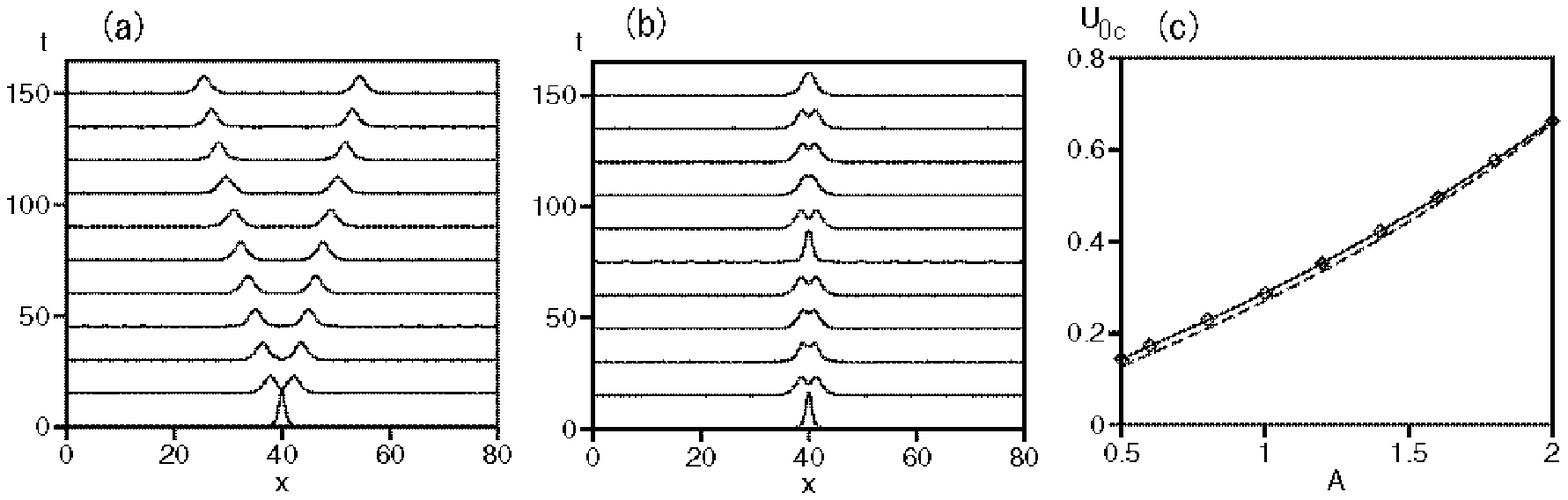}
\end{center}
\caption{(a) Time evolution of $|\phi(x,t)|$ for $U_0=0.67,A=2$ and $L=80$. A soliton is splitted into two pulses by the potential wall of $d=1$. (b) Time evolution of $|\phi(x,t)|$ for $U_0=0.66,A=2$ and $L=80$. A soliton keeps its one-pulse structure, although it is breathing. (c) Critical value of $U_0$ for the splitting of the soliton as a function of the initial amplitude $A$. The dashed line is the theoretical one $U_0=A^2/\{8\tanh(Ad/2)\}$}
\label{f2}
\end{figure}
The transmission coefficients as shown in Fig.~1(b) change more and more sharply as $U_0$ is decreased. As $U_0$ becomes smaller, such a splitting or additional radiations are hardly observed and the soliton keeps its form and behaves like a particle.  The coefficient jumps almost from 0 to 1 at a certain critical velocity for $U_0<0.05$.  To understand the transition between the wavy behaviors including the splitting and the particle-like behavior, we have performed numerical simulations for another initial condition $\phi(x,0)=A{\rm sech}\{A(x-L/2)\}$, that is, the soliton is initially set at $L/2$ and the initial velocity is 0. The potential wall is located in $[L/2-d/2,L/2+d/2]$. The mirror-symmetry around $L/2$ is maintained during the time evolution.   For large $U_0$, the soliton is split into two pulses as shown in Fig.~2(a) for $U_0=0.67, A=2,d=1$ and $L=80$ owing to the repulsive potential.  However, the soliton cannot be split for small $U_0$ owing to the attractive interaction by the nonlinear term. The soliton keeps its one pulse structure  as shown in Fig.~2(b) for $U_0=0.66, A=2,d=1$ and $L=80$. The critical value of $U_0$ depends on the initial amplitude $A$ as shown in Fig.~2(c). Equation (1) has at least two invariants, that is, the total mass $N=\int_0^L|\phi|^2dx$ and the total energy $E=\int_0^L\{1/2|\phi_x|^2-1/2|\phi|^4+U(x)|\phi|^2\}dx$.  If the soliton is split into two counterpropagating solitons: $\phi(x,t)=A^{\prime}{\rm sech}\{A^{\prime}(x-vt-L/2)\}e^{ip(x-vt-L/2)-i\omega t}+A^{\prime}{\rm sech}\{A^{\prime}(x+vt-L/2)\}e^{-ip(x+vt-L/2)-i\omega t}$, the amplitude $A^{\prime}$ and the momentum $p$ satisfy $A^{\prime}=A/2$ and $2p^2A^{\prime}=-1/4A^3+2U_0A\tanh(Ad/2)$
 owing to the conservation laws. If $U_0<A^2/\{8\tanh(Ad/2)\}$, there is no real solution for $p$, that is, the splitting cannot occur and the one-pulse structure is maintained.  Above the line $U_0=A^2/\{8\tanh(Ad/2)\}$, the potential energy is used to split a soliton against the attractive interaction. A soliton behaves like a particle even for more general initial conditions, if $U_0$ is sufficiently smaller than  $A^2/\{8\tanh(Ad/2)\}$.  

We focus on the particle-like behavior below. We return to the numerical simulations of the first type. Figures 3(a) and (b) display the time evolutions of a soliton passing through a potential wall of $A=1,U_0=0.04$ and $d=1$. The system size is 50 and the potential wall locates at $x_0=L/2=25$. The soliton is almost completely reflected when the initial velocity is $v=0.18$, and almost completely transmitted for $v=0.2$.  The critical velocity for the soliton to pass through the
\begin{figure}[t]
\begin{center}
\includegraphics[width=14cm]{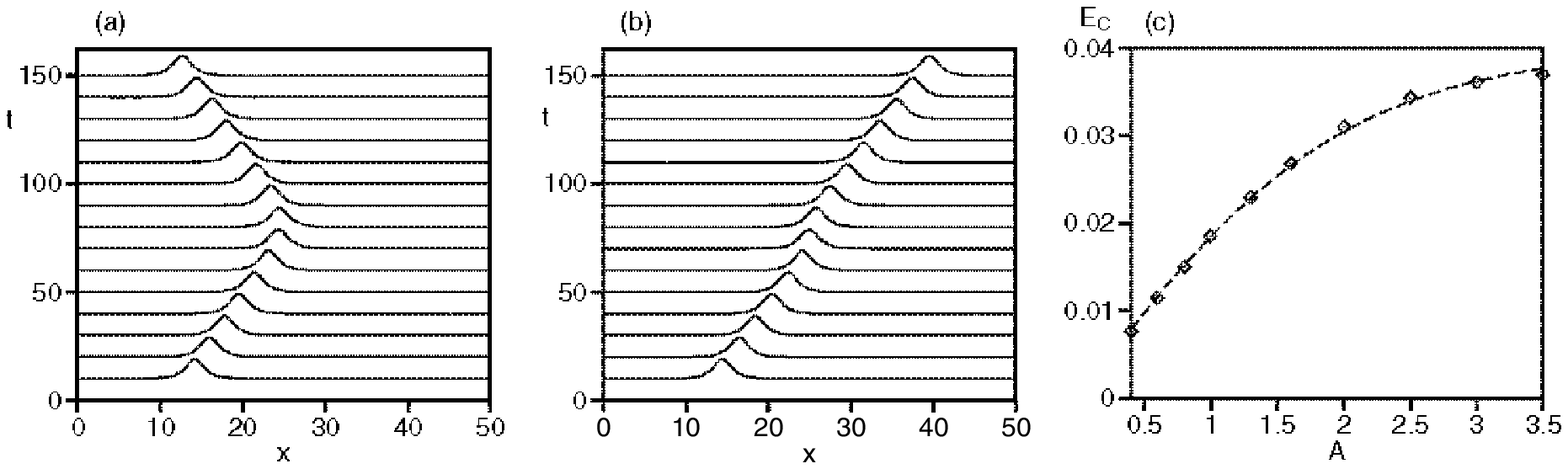}
\end{center}
\caption{(a) Time evolution of $|\phi(x,t)|$ for $U_0=0.04,A=1$ and $v=0.18$. (b) Time evolution of $|\phi(x,t)|$ for $U_0=0.04,A=1$ and $v=0.2$. (c) Critical kinetic energy in the potential wall of $U_0=0.04$ as a function of $A$.
The dashed curve is a theoretical approximation $0.04 \tanh(A/2)$}
\label{f3}
\end{figure}
 potential wall is 0.1925, and the critical kinetic energy is 0.0185, which is smaller than the critical value $E_c=U_0=0.04$ for a particle obeying the classical mechanics. Figure 3(c) displays the critical kinetic energy as a function of the amplitude $A$ of the soliton. As the amplitude $A$ is increased, the critical energy approaches $U_0$. A soliton with smaller amplitude can pass easily through the potential wall.  This may be interpreted as coherent tunneling. 

The nonlinear Schr\"odinger equation is rewritten with the Lagrangian form: $\partial/\partial t(\delta L/\delta \phi_t)=\delta L/\delta \phi$,
where $L=\int_{-\infty}^{\infty} dx \{i/2(\phi_t\phi^*-\phi_t^*\phi)-1/2|\phi_x|^2+1/2|\phi|^4-U(x)|\phi|^2\}$.
If a solution to the nonlinear Schr\"odinger equation is approximated by  
\begin{equation}
\phi(x,t)=A(t){\rm sech}\{(x-\xi(t))/a(t)\}e^{ip(t)(x-\xi(t))-i\omega t},
\end{equation}
the time evolution of $a(t),\xi(t)$ and $p(t)$ can be derived from the effective Lagrangian $L_{eff}=\int_{-\infty}^{\infty}L dx$~\cite{Trom} using 
\[L_{eff}=2A^2ap\xi_t+2/3A^4a-1/3A^2/a-A^2ap^2-2A^2aU_{eff},\]
where $U_{eff}=U_0/2[\tanh\{(x_0+d-\xi)/a\}-\tanh\{(x_0-\xi)/a\}].$
Equations of motion for the variational parameters $q_i=\xi,p$ and $a$ are given by $d/dt(\partial L_{eff}/\partial \dot{q_i})=\partial L_{eff}/\partial q_i$ under the condition $N=\int_{-\infty}^{\infty}|\phi|^2dx=2A^2a=const$. The results are  
\begin{equation}
\frac{d\xi}{dt}=p,\;\;\frac{dp}{dt}=-\frac{\partial U_{eff}}{\partial \xi},\;\;\frac{da}{dt}=0
\end{equation}
and $A=1/a=N/2$. The soliton obeys the Newton equation with an effective potential $U_{eff}$. The peak amplitude of the effective potential is $U_0\tanh(Ad/2)$. The critical kinetic energy is therefore expected to be $E_c=U_0\tanh(Ad/2)$, which was drawn with the dashed curve in Fig.~3(c).  Good agreement of numerical results with this approximation is seen in Fig.~3(c). 

\begin{figure}[t]
\begin{center}
\includegraphics[width=12cm]{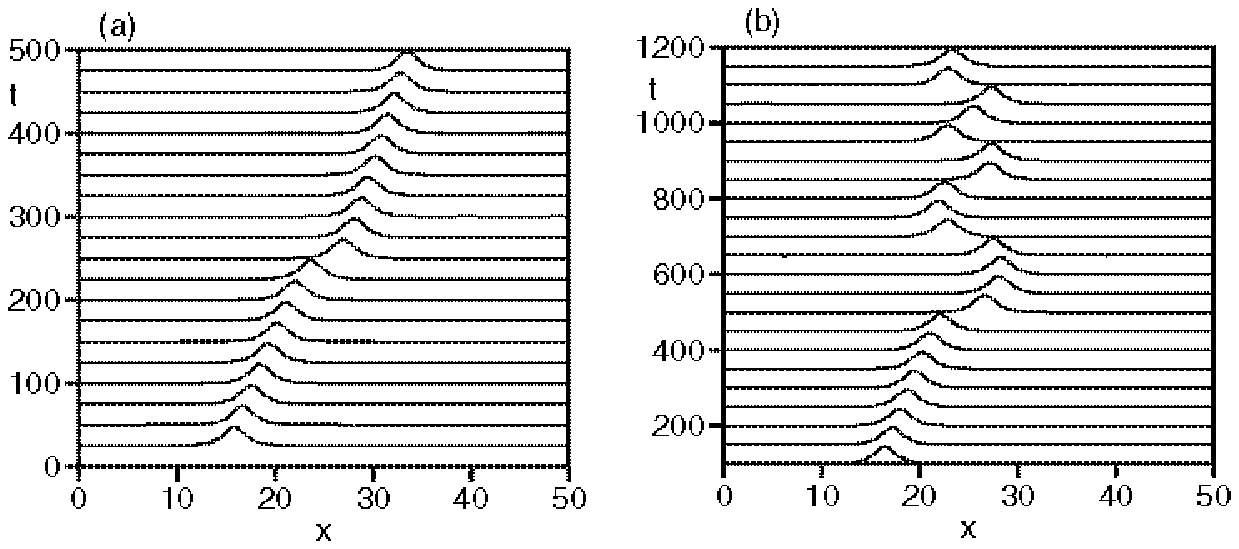}
\end{center}
\caption{Time evolutions of $|\phi(x,t)|$ in a potential well of $U_0=-0.04$ and $d=1$. The initial velocity is  0.035 (a) and 0.015 (b).}
\label{f4}
\end{figure}
Secondary, we consider the case of negative $U_0$. The potential acts as a potential well.  We study the scattering of a soliton by the potential well. 
Complicated scattering including splitting is observed when $|U_0|$ is large, as is the case of positive $U_0$. We focus on the particle-like behavior, which is observed when $|U_0|$ is sufficiently small. 
Figure 4 displays the time evolutions of solitons for a potential well with $U_0=-0.04$ and width $d=1$. The initial amplitude is $A=1$ and the initial velocities are $v=0.035$ and 0.015.  For relatively large $v$, the soliton passes over the potential well like a classical mechanical particle obeying eq.~(6). The velocity increases when the particle passes over the potential well, but its initial value is recovered after the passage. However, the soliton is trapped by the potential well for sufficiently small $v$ as shown in Fig.~4(b). The critical value of $v$ for the trapping is approximately $v_c\sim 0.0225$. In Fig.~4(a), a soliton passes through the potential well, but the velocity is decreased from 0.035 to 0.026 after the passage of the potential well. Some dissipative processes may occur at the potential well. Some small radiations may be generated and a part of the kinetic energy is changed into radiation energy. Another possibility is that some internal freedoms of a soliton such as breathing motion may be excited and the kinetic energy decreases. To investigate such motion, we assume a solution of the form  
\begin{equation}
\phi(x,t)=A(t){\rm sech}\{(x-\xi(t))/a(t)\}e^{ip(t)(x-\xi(t))+i\sigma(t)\log \cosh\{(x-\xi(t))/a(t)\}-i\omega t},\end{equation}
where $A,a,p,\xi$ and $\sigma$ change in time, but $N=2A^2a$ is conserved in time. The term  $i\sigma(t)\log \cosh(x-\xi(t))/a(t)$ is a kind of chirp term and this form of chirp term appears in a solution to the complex Ginzburg-Landau equation where some dissipation and filtering effects are involved.  Owing to the inclusion of this term, the width of the soliton changes in time and breathing motion becomes possible. The choice of this form of chirp term is not a unique one. We have studied another form of chirp term $i\sigma(x-\xi(t))^2$, but better agreement with direct numerical simulation was seen for the form eq.~(7).  The substitution of eq.~(7) to the Lagrangian and its spatial integration yield the effective Lagrangian
\[L_{eff}=N(p\xi_t-\sigma_t(1-\log 2)+\sigma a_t/(2a)-1/(6a^2)-p^2/2-\sigma^2/(6a^2)+N/(6a)-U_{eff}),\]
where $U_{eff}=U_0/2[\tanh\{(x_0+d-\xi)/a\}-\tanh\{(x_0-\xi)/a\}].$ 
The equations of motions for $p(t),\xi(t),a(t)$ and $\sigma(t)$ are 
\begin{eqnarray}
\frac{d\xi}{dt}&=&p,\;\;
\frac{dp}{dt}=-\frac{\partial U_{eff}}{\partial \xi},\nonumber\\
\frac{da}{dt}&=&\frac{2\sigma}{3a},\;\;
\frac{d\sigma}{dt}=2a\left(\frac{1+\sigma^2}{3a^3}-\frac{N}{6a^2}-\frac{\partial U_{eff}}{\partial a}\right ).
\end{eqnarray}
The energy $p^2/2+U_{eff}$ is not conserved in this equation, and the width $a$ changes in time. If the energy decays, the soliton may be trapped in the potential well. 
\begin{figure}[t]
\begin{center}
\includegraphics[width=11cm]{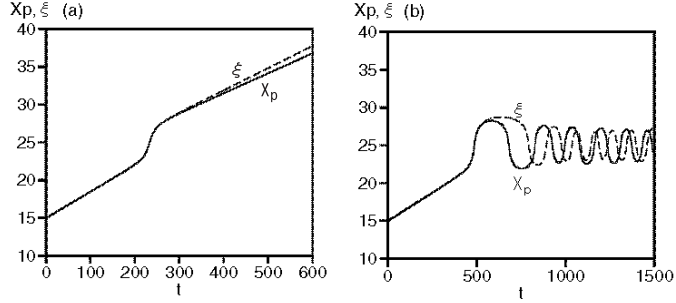}
\end{center}
\caption{Comparison of the time evolutions of the peak position $X_p(t)$ of a soliton (solid line) and $\xi(t)$ by Eq.~(8) (dashed line) for $v=0.035$ (a) and 0.015 (b).}
\label{f5}
\end{figure}
Figure 5 displays the time evolutions of $\xi(t)$ by eq.~(8) for $v=0.035$ and $v=0.015$ with the dashed curves,  and they are compared with the time evolutions of the peak position of the solitons shown in Figs.~4(a) and (b). The velocity of a particle changes from the initial value of 0.035 to 0.029 in the time evolution of eq.~(8) in Fig.~5(a). The particle is trapped in the potential well and is oscillating in a region of $23<x<28$ for the initial velocity $v=0.015$ in Fig.~5(b).  The trapping occurs for $v<0.019$ in eq.~(8). The dissipative effect seems to be slightly weaker in eq.~(8) than in the nonlinear Schr\"odinger equation eq.~(1). 

\begin{figure}[t]
\begin{center}
\includegraphics[width=12cm]{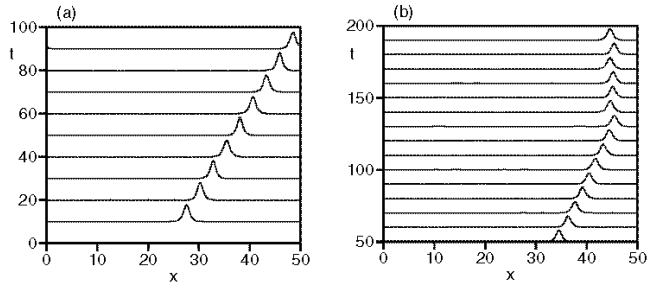}
\end{center}
\caption{Time evolutions of $|\phi(x,t)|$ in a periodic potential $U(x)=-0.2\cos(2\pi x)$. The initial amplitude $A$ is 2, and the initial velocity is $v=0.33$ (a) and 0.28 (b).}
\label{f6}
\end{figure}
As the third case, we study soliton motion in a spatially periodic potential $U(x)=-U_0\cos(2\pi x)$. Initially, a soliton locates at the position of the potential minimum $x=L/2=25$, and the amplitude is $A=2$. Figure 6 displays the time evolutions of solitons in the periodic potential with $U_0=0.2$ for two different initial velocities $v$. A soliton is almost steadily propagating for $v=0.33$ as shown in Fig.~6(a). For $v=0.28$, a soliton propagates in an initial stage, however, it is trapped in the periodic potential  as shown in Fig.~6(b). 
The critical velocity for a soliton to be trapped for $A=2$ is approximately 0.312.
\begin{figure}[t]
\begin{center}
\includegraphics[width=11cm]{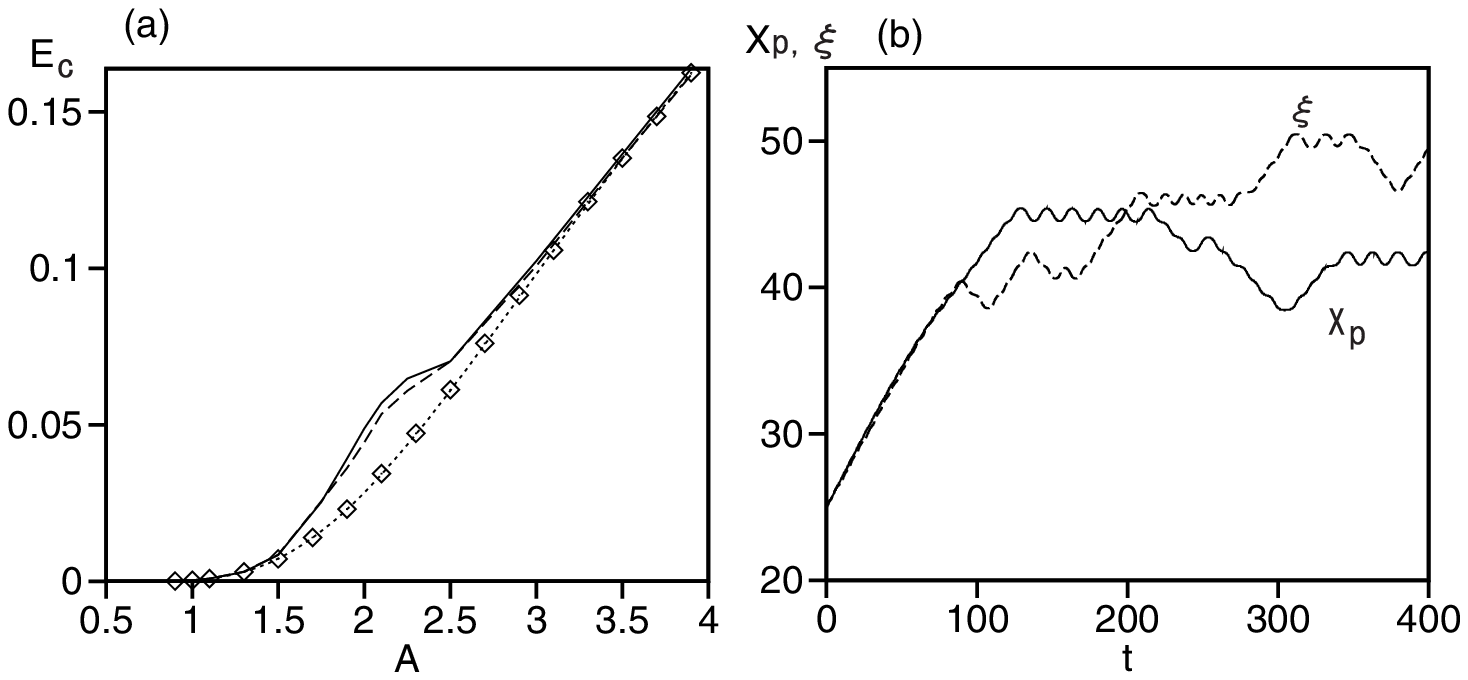}
\end{center}
\caption{(a) Critical kinetic energy for the propagation in a periodic potential
 $U(x)=-0.2\cos(2\pi x)$ as a function of $A$. The solid line denotes the critical kinetic energy, above which a soliton propagates almost steadily. 
The rhombi denote critical values, below which a soliton cannot go over the nearest potential hump. The dotted curve is a theoretical line  $0.4\pi^2/(A\sinh(\pi^2/A))$. The dashed line denotes a critical kinetic energy above which a particle obeying eq.~(8) propagates almost steadily. (b) Time evolutions of the peak position $X_p(t)$ by eq.~(1) (solid line) and $\xi(t)$ by eq.~(8) (dashed line).}
\label{f7}
\end{figure}
Figure 7(a) displays the critical kinetic energy $E_c=v_c^2/2$ as a function of $A$. The solid curve represents the critical kinetic energy, and the rhombi represent critical points below which a soliton cannot go over the nearest potential hump. (For the parameter of $v$ shown in Fig.~6(b), the soliton goes over several potential humps and then it is trapped.) The effective potential $U_{eff}$ for the periodic potential is expressed as 
\[U_{eff}=\int_{-\infty}^{\infty}\{-U_0\cos(2\pi x)\}|\phi|^2dx/N=\frac{-U_0\cos(2\pi \xi)\pi^2a}{\sinh(\pi^2 a)}.\]
A particle, which obeys eq.~(6) with the effective potential $U_{eff}$ and starts from a position of potential minimum, exhibits a
 transition from a trapped state to a propagating state. 
The critical initial kinetic energy  is $2U_0\pi^2/(A\sinh(\pi^2/A))$, which is drawn by the dotted curve in Fig.~7(a). The critical curve is a good approximation for the critical values marked with rhombi. 
 This is probably because eq.~(6) is a good approximation to the short-time evolution of eq.~(1), where dissipative effects are negligible, and which determines whether the soliton can pass over the first neighboring hump.  In the region $1.5<A<3$, the critical curve $2U_0\pi^2/(A\sinh(\pi^2/A))$ is rather different from the true critical curve denoted by solid line. If the width of a soliton $2/A$ is comparable to the wavelength of the periodic potential, the interaction between the soliton and the potential will be enhanced, which may be an origin of the deviation. The trapping phenomena accompanying the dissipation of kinetic energy can be expressed with eq.~(8) with $U_{eff}=-U_0\cos(2\pi \xi)\pi^2a/\sinh(\pi^2 a)$. The time evolution of $\xi(t)$ by eq.~(8) for $A=2$ and $v=0.2785$ is displayed in Fig.~7(b) with the dashed curve. The solid curve in Fig.~7(b) is the time evolution of the peak position $X_p(t)$ of the soliton for $A=2$ and $v=0.28$. Similar time evolutions including damping of kinetic energy, trapping and chaotic motions are observed for $\xi(t)$ and $X_p(t)$, although the initial velocities are slightly different. Since the number of degrees of freedom in eq.~(8) is four, it is not surprising that chaotic motion appears in eq.~(8). The dashed curve in Fig.~7(a) is the critical line, above which steadily propagating solutions appear in eq.~(8).  There is fairly good agreement between the solid line and the dashed line, although the dashed line locates slightly below the sold line in the region $1.5<A<3$. 

In summary, we have performed numerical simulations for the scattering and the trapping of nonlinear Schr\"odinger solitons in some external potentials.
The transition from the reflection to the transmission by a potential wall 
depends strongly on the soliton amplitude. We have also found the trapping of solitons in a potential well and a periodic potential. The critical energy for the transitions can be approximately evaluated using eq.~(6) or (8) derived from a variational method.  


\begin{thebibliography}{9}
\bibitem{Karp} V.~I.~Karpman: {\it Nonlinear Waves in Dispersive Media} (Pergamon Press, Oxford, 1975).
\bibitem{Hase} A.~Hasegawa: {\it Optical Solitons in Fibers} (Springer-Verlag, Berlin, 1990). 
\bibitem{Khay} L.~Khaykovich et al.: Science {\bf 296} (2002) 1290.
\bibitem{Moura} M.~de Moura: J. Phys. A {\bf 27} (1994) 7157.
\bibitem{Frau} H. Frauenkron and P.~Grassberger: Phys. Rev. E {\bf 53} (1996) 2823.
\bibitem{Baiz} B.~B.~Baizakov, V.~V.~Konotop and M.~Salerno: J.~Phys. B {\bf 35} (2002) 5105.
\bibitem{Trom} A.~Trombettoni and A.~Smerzi: Phys. Rev. Lett. {\bf 86} (2001) 2353. 
\end{thebibliography}
\end{document}